\documentclass[aps,amsmath,twocolumn,floatfix,showpacs,superscriptaddress,footinbib,longbibliography]{revtex4-1}
\pdfoutput=1
\usepackage[dvips]{graphics}
\usepackage{bm}
\usepackage{import}
\usepackage{epsfig}
\usepackage{enumerate}
\usepackage{subfigure}
\usepackage{color}
\usepackage{xcolor}
\usepackage{epsfig,amssymb,amsmath,latexsym}
\usepackage{sidecap,tikz}
\usepackage{graphicx}
\usepackage{makecell}
\usepackage{hyperref}
\usepackage{multirow}
\usepackage{bm}
\usepackage{MnSymbol}
\usepackage{braket}
\usepackage{float}
\usepackage{subfigure}
\usepackage{tikz}

\usepackage[normalem]{ulem}

\hypersetup{
    pdfnewwindow=true,      
    colorlinks=true,       
    linkcolor=blue,          
    citecolor=magenta,        
    filecolor=blue,      
    urlcolor=blue        
}

\newcommand{\etal}{{\it et al.~}}
\newcommand{\ie}{{i.e., }}

\newcommand{\bea}{\begin{eqnarray}}
\newcommand{\eea}{\end{eqnarray}}
\newcommand{\beq}{\begin{equation}}  
\newcommand{\eeq}{\end{equation}}

\newcommand\mb{\mathbf}
\newcommand{\sg}{\sigma}

\newcommand{\UP}{\uparrow}
\newcommand{\DN}{\downarrow}

\newcommand{\mbf}{\mathbf}
\newcommand{\mc}{\mathcal}

\definecolor{lime}{HTML}{A6CE39}
\DeclareRobustCommand{\orcidicon}{\hspace{-1mm}
	\begin{tikzpicture}
	\draw[lime, fill=lime] (0,0) 
	circle [radius=0.16] 
	node[white] {{\fontfamily{qag}\selectfont \tiny \,ID}};
	\draw[white, fill=white] (-0.0525,0.095) 
	circle [radius=0.007];
	\end{tikzpicture}
	\hspace{-3mm}
}
\foreach \x in {A, ..., Z}{\expandafter\xdef\csname orcid\x\endcsname{\noexpand\href{https://orcid.org/\csname orcidauthor\x\endcsname}
			{\noexpand\orcidicon}}
}

\begin{document} 
\title{Thermal transport in superconductor heterostructures: some recent progress}
\author{Paramita Dutta\orcidA{}}
\affiliation{Theoretical Physics Division, Physical Research Laboratory, Navrangpura, Ahmedabad-380009, India}
\email{paramita@prl.res.in}
\date \today

\begin{abstract}
This article reviews recent advances in low-temperature electronic thermal transport properties of thermally biased superconductor heterostructures focusing on the two-terminal transport. Since the last decade, ferromagnetism has been widely used to enhance the thermoelectricity in heterostructures based on ordinary superconductors. The possibility of getting giant thermoelectric effects with optimum thermal conductance by breaking the electron-hole symmetry of the ordinary superconductor boosted the research in this direction. Recently, attention has been paid to the role of triplet Cooper pairs that emerged in ferromagnetic junctions and the possibility of advanced applications. Other forms of magnetism, specifically antiferromagnetism and altermagnetism, have been investigated to unravel the behavior of the thermal and charge current in thermally biased junctions. In parallel to ordinary superconductors, junctions with unconventional superconductors have been explored for the same purpose. Thermal transport in superconducting bilayers has been studied using advanced materials like Dirac and topological materials, including Weyl semimetals. Significant attention has been paid to thermally biased topological Josephson junctions to explore the phase-tunable current in recent times. Weyl Josephson junctions, multi-terminal Josephson junctions, and various other multilayer junctions have also been studied to engineer large thermoelectric effects and various functionalities with potential applications in superconductor-based thermal device components.

\end{abstract}

 \maketitle

\tableofcontents

\section{Introduction}

The study of thermal transport properties is one of the powerful ways to characterize properties of condensed matter systems. It can bring additional information that cannot be found by investigating the electrical transport phenomena. Investigating the effects of biasing the system by temperature gradient and understanding the origin and implications of thermal transport is of great interest to researchers from the perspectives of both fundamental physics and applications in heat management and devices~\cite{Goldsmid2017,Goldsmid1995,Giazotto2006}. 

At low temperatures (sub-$100$mK), electrons get thermally decoupled from the lattice in general~\cite{Giazotto2006}. Thus, phonon contributions can be ignored to separate the electronic part at this regime. There are two major aspects of the electronic thermal transport phenomena in a thermally biased systems: the generation and behaviors of the (i) charge current without any bias voltage when the heat is converted into electricity, i.e., the so-called thermoelectricity, and (ii) the heat current due to the applied thermal gradient. The effect of the simultaneous presence of voltage and thermal bias is also studied. 

To study thermal transport phenomena, normal metals (NMs) were initially considered. Superconductors were included almost hundred years ago~\cite{Ginzburg1944,Galperin1974,Ginzburg1978,Ginzburg1989,Andreev1,Clarke1979}. However, the seminal work by Ginzburg describing distinctive thermoelectric effects in superconductors using two-fluid model appeared in 1944~\cite{Ginzburg1944}. It was followed by measurements in bi-metallic superconducting loops~\cite{Zavaritskii1974,Falco1976,Harlingen1980}.
Thermoelectricity in the bulk of gapped superconductors was predicted to be low compared to NMs at the very beginning, because of the opposite flows of normal excitation and Cooper pair condensate current. The separation of two charge currents became essential to realize the thermoelectric effects in it~\cite{Galperin1974}. Later, the contributions from the convection flux originated from the rupture of Cooper pairs at the higher temperature and the conversion of the normal electrons to the Cooper pairs at the lower temperature were identified~\cite{Ginzburg1989}. Thus, unlike NM conductors, in superconductors the dominating process is decided by the system temperature. However, the symmetry of the gap in the superconductor density of states, and the interference of two charge currents generated due to the thermal bias result in low or even vanishing thermoelectric current.

To intensify the thermoelectric effect, the concept of adding inhomogeneity or impurity in superconductors that can break the symmetry of the density of states, was then introduced~\cite{Ginzburg1978}. The impurity scattering can lead to larger current due to the formation of quasibound Andreev states. This yields high asymmetry between the electron and hole scattering rate, leading to a giant thermoelectric effect~\cite{Kalenkov2012}. To enhance the thermoelectricity, superconducting hybrid junctions are effective solutions because of the strategy of symmetry breaking either by using ferromagnet (FM), magnetic impurity, spin-orbit coupling, or advanced quantum materials attached to even fully gapped ordinary superconductor(s)~\cite{Kalenkov2012,Machon2013,Machon2014,Ozaeta2014,Giazotto2015,Kolenda2016a,Kolenda2016b,Hwang2016,Linder2016,Kolenda2017,Dutta2017,Sothmann2018a,Beckmann2019,Heikkila2019,Seviour2000,Dutta2020b,Ouassou2022,Geng2023,Blasi2020a,Blasi2020b,Blasi2021,Hwang2020,Beiranvand2017,Shapiro2017,Bours2018,Bours2019,Scharf2020,Scharf2021,Mukhopadhyay2021,Dutta2023,Chatterjee2024}. These hetrostructures have drawn a lot of attention especially after the experimental verification of the thermopower in FM/superconductor junction in 2016~\cite{Kolenda2016a}, where an agreement with the theoretical prediction~\cite{Ozaeta2014} was reached. The heat calibration measurements with Seebeck coefficient $\sim\,100\mu$V/K in high field superconductor/FM tunnel junction became a landmark on the roadmap of the thermoelectric effect in superconductor heterostructures~\cite{Kolenda2016a}. For comparison, note that, NM shows up to $10 \mu $V/K thermopower at room temperatures~\cite{Rowe1995}.

In ferromagnetic junctions, a three-terminal setup utilizing the nonlocal crossed Andreev reflection and elastic cotunneling processes~\cite{Machon2013}, four-terminal setup~\cite{Wu2021}, spin splitting, and spin filtering effects for significant accumulation of quasiparticle density of states at energies near the superconducting gap edge~\cite{Machon2014,Ozaeta2014}, quasiparticle tunneling between two spin-split superconductors~\cite{Linder2016}, a combination of spin-polarized tunneling at ferromagnetic–quantum dot interface~\cite{Hwang2016,Kamp2019}, ferromagnetic insulator-based superconducting tunnel junctions~\cite{Giazotto2020} have been utilized to boost the thermoelectric effect in them. The theoretical prediction was for the Seebeck coefficient of the order of mV with the huge thermoelectric figure of merit ($zT\sim 40$)~\cite{Linder2016}. Current rectification in junction using spin-split superconductors~\cite{Illic2022}, thermally induced spin-transfer torque~\cite{Bobkova2021}, temperature-dependent spin transport and current-induced torques~\cite{Muller2021}, thermoelectric generation of equal-spin Cooper pairs~\cite{Keidel2020} have been explored in ferromagnetic junctions. For the theoretical calculations, quasiclassical formalism~\cite{Kalenkov2012,Machon2013,Machon2014}, Boltzmann equation description~\cite{Aronov1981}, tunneling Hamiltonian calculations~\cite{Ozaeta2014} have been adopted. From the application perspective, Al film~\cite{Giazotto2015}, Fe~\cite{Kolenda2016a}, NbN/GdN/NbN~\cite{Linder2016} are predicted to be useful materials. 

In addition to ferromagnetism, antiferromagnetism is also used to explore the thermal transport~\cite{Jakobsen2020}. Thermally induced spin torque and domain-wall motion were studied in antiferromagnetic insulator based superconducting bilayer~\cite{Bobkov2021}. This tradition of adding magnetism has been recently extended to altermagnetism very recently~\cite{Sukhachov2024,Chourasia2024}. Apart from adding a form of magnetism or magnetic impurity, advanced materials like Dirac materials e.g. graphene~\cite{Yokoyama2008,Beiranvand2017,Paul2016,Aydin2022,Huang2023}, silicene~\cite{Guzman2018}, bilayer graphene~\cite{Di2021,Bernazzani2023,Bera2025}, topological insulator (TI)~\cite{Ren2013}, Weyl semimetal (WSM)~\cite{Saxena2023, Chatterjee2024} have been used to modify the currents in thermally biased superconducting heterostructures. The advancement in the fabrication technique over the past decades have stimulated the investigation of thermal transport at the nanoscale level with enhanced potential for the application in quantum devices based on electronic solid-state cooling, thermal switching effect, thermal detector, quantum sensing etc.~\cite{Giazotto2006}. 

The search for ways to control the thermoelectricity in superconductor bilayers has been continued by adding another superconductor layer, forming Josephson junctions. Attention has been paid to the Josephson junctions since the phase tunability of the current by external flux offers an additional freedom here. A non-trivial thermal bias-induced voltage can be generated by tuning the superconducting phase of the junction~\cite{Guttman1997,Kalenkov2020,Marchegiani2020,Blasi2020a,Blasi2020b,Mukhopadhyay2022}. Similar to the bilayer junctions, various materials have been considered for Josephson junctions too ~\cite{Giazotto2015,Huang2023,Chatterjee2024}, but special attention has been paid to topological materials to generate the phase-tunable thermal current and to detect topological bound states via thermal current and other functionalities~\cite{Wiedenmann2016,Sothmann2016a,Sothmann2017,Bours2018,Bours2019,Blasi2020a,Blasi2020b,Blasi2021,Hwang2020,Scharf2020,Scharf2021,Gresta2021,Saxena2022,Mukhopadhyay2022,Dutta2023} as the combination of global topology and local superconducting order has been established to host exotic transport properties in the literature~\cite{Fu2008,Fu2009,Hasan2010,Qi2011,Black-Schaffer2012,Black-Schaffer2013,Cayao2017,Dutta2019,Dutta2020a,Dutta2024,Sten2025}. The major advantage of using the topological property is that one can avoid using external magnetic components like magnetic impurity~\cite{Kalenkov2012} or creating any vacancy~\cite{Aydin2022} to manipulate the current but side effects like the existence of stray fields cannot be avoided. Depending on the topology and thermal gradient, a sizeable phase-coherent charge current is shown to be generated in the topological Josephson junction~\cite{Kalenkov2020}. Junctions involving TIs have drawn significant attention because of their potential to get rid of the backscattering~\cite{Tanaka2004,Rolf2013,Islam2017,Cayao2022} and, most importantly, to host Majorana fermions, which are predicted to be helpful for fault-tolerant quantum computation~\cite{Kane2005,Fu2008,Fu2009,Qi2011,Tkachov2013}.

Other works on thermal transport in Josephson junctions involve the study of the Josephson heat interferometer~\cite{Giazotto2012}, signatures of topological Andreev bound states (ABSs) in phase-dependent heat transport~\cite{Sothmann2016a}, phase-coherent heat circulators using multiterminal Josephson junctions~\cite{Sothmann2018b}, phase-tunable thermal rectifiers~\cite{Martinez2013,Chatterjee2024}, thermal current in the nonlinear regime in a spontaneously broken particle-hole symmetry system\,\cite{Marchegiani2020},
thermal noise effect in ferromagnetic Josephson junctions~\cite{Guarcello2021}, thermospin effect induced by spontaneous symmetry breaking in superconducting tunnel junctions~\cite{Germanese2022}, and recently,
thermal superconducting quantum interference transistor (T-SQUIPT)~\cite{Ligato2022}. Thermal transport study has been extended to other multilayers of superconductor heterostructures too~\cite{Sothmann2021,Ouassou2022,Araujo2024,Battisti2024,Chen2025,Tuero2025}. In contrast to using symmetry-breaking non-superconducting elements, hybrid junctions using unconventional superconductors have also been shown to be useful~\cite{Devyatov2000,Savander2020,Guarcello2023,Pal2024b,Klees2024,Trocha2025,Matsushita2025}. 

With this state-of-the-art, we now present the recent advancements in this sub-field. We start with a discussion on bilayers based on FM followed by antiferromagnet (AFM), altermagnet (AM), other advanced materials like Dirac materials, TI, WSM, etc., and also bilayer using unconventional superconductors in Sec.\,\ref{Sec:bilayer}.  
Then, we illustrate the progress in Josephson junctions and other multilayers in Sec.\,\ref{Sec:JJ} and Sec.\,\ref{multi}, respectively. Throughout this review, we avoid describing the theoretical formalism in detail except for model Hamiltonians and some expressions used for fundamental calculation for thermal transport, and focus on recent results. Finally, we summarize and conclude in Sec.\,\ref{Sec:Summary}.

\section{Bilayer heterostructures} \label{Sec:bilayer}

In this section, we discuss the advancements in the research on the thermoelectricity in bilayer junctions where a superconductor is attached to a non-superconducting material with a temperature gradient applied across the junction. We start by describing some fundmentals about the theoretical formalism before we discuss various bilayer models and results in different subsections. 
 
The applied temperature gradient affects the junction in two different ways: (i) it modifies the superconducting gap and (ii) it affects the occupation number of the quasiparticles within the superconductor~\cite{Pershoguba2019}. The temperature gradient induces two different currents in the junction: (i) charge current and (ii) heat current. Now the charge current again may posses (i) dissipative and (ii) non-dissipative parts. The temperature gradient is considered as sufficiently low so that the transport phenomena are described within the linear regime unless specified. 

\subsection{Theoretical formalism}

For the theoretical description of bilayer heterostructures, Bogoliubov-de Gennes (BdG) Hamiltonian is employed as,
\beq
H_{\rm{BdG}}= \frac{1}{2} \sum_k \Psi_{\bf{k}}^\dagger \mc{H}({\bf{k}}) \Psi_{\bf{k}}
\eeq
where $\Psi_\mbf{k}=[c_{\mbf{k},\UP},c_{\mbf{k},\DN},c^\dagger_{-\mbf{k},\DN},-c^\dagger_{-\mbf{k},\UP}]^T$ with $c_{\mbf{k}\sg}~(c_{\mbf{k}\sg}^\dagger)$ as the electron annihilation (creation) operator for the momentum $\mbf{k}\in \{k_x,k_y\}$ and spin $\sg$, and $\mc{H}(\mbf{k})$ the first quantized Hamiltonian. 

Within the linear response regime, the charge and heat currents are related to the bias voltage $\Delta V$ and temperature gradient $\Delta T$ following the Onsager relation given by~\cite{Onsager1931a,Onsager1931b}, 
\begin{equation}
\begin{bmatrix}
	~ I_c \\I_q 
\end{bmatrix} = \begin{bmatrix}
~ \mc{L}_{11}  & \mc{L}_{12} \\
~ \mc{L}_{21} & \mc{L}_{22} 
\end{bmatrix}  \begin{bmatrix}
	~ \Delta V \\ \Delta T 
\end{bmatrix}\ ,
\end{equation}
where $\mc{L}_{11}$ and $\mc{L}_{22}$ are the electrical and thermal conductance, respectively, describing the conjugate processes. The off-diagonal elements represent the non-conjugate processes where charge (heat) current flow due to the temperature gradient (voltage bias) \ie $\mc{L}_{12}$ ($\mc{L}_{21}$). The thermoelectric coefficients expressed as~\cite{Riedel1993,Wysokinski2012,Dutta2017,Dutta2020b,Pal2024b},
\begin{eqnarray}
\mc{L}_{11}&=& \frac{e^2}{h} \int_{0}^{\infty} dE \,\mc{T}(E) \left(-\frac{\partial f (E,T)}{\partial E}\right) \label{Eqn. L11},	\\
\mc{L}_{12}&=& \frac{e}{hT}\int_{0}^{\infty} dE \,\mc{T}(E) E \left(-\frac{\partial f (E,T)}{\partial E}\right),  \label{Eqn. L12} \\
\mc{L}_{22}&=& \frac{1}{hT} \int_{0}^{\infty} dE \,\mc{T}(E) E^2 \left(-\frac{\partial f (E,T)}{\partial E}\right)  ,
\label{Eqn. L22}
\end{eqnarray}
where $\mc{T}(E)$ is the transmission function of carriers. It governs the thermoelectric response primarily. For superconductor bilayer junctions, these transmission functions are expressed in terms of the ordinary and Andreev reflections~\cite{Riedel1993,Wysokinski2012,Dutta2017,Dutta2020b,Pal2024b}. BTK formalism~\cite{Blonder1982} has been extensively used to calculate the transport properties. 

From the application perspectives, thermoelectricity is studied in terms of the Seebeck effect (also known as thermopower) which measures the voltage developed in the system per unit temperature gradient in the open-circuit condition. The Seebeck coefficient $\mc{S}$, in units of $k_B/e$, is defined as,
\bea
	\mc{S}=-\frac{\Delta V}{\Delta T} = \frac{\mc{L}_{12}}{\mc{L}_{11}} \label{Eq. Seebeck}
	\eea
and the efficiency is described in terms of the dimensionless quantity, thermoelectric figure of merit, $zT$ given by,
	\bea
	zT=\frac{\mc{S}^2 \mc{L}_{11}T}{\mc{L}_{22}-\mc{L}_{12}^2/(\mc{L}_{11}T)} \label{Eq. zT}\
\eea
where $T$ is the base temperature of the system. 

Before we proceed to discussing recent results for various junctions, a few comments are in order. The temperature-dependance of the superconducting pair correlation is expressed as $\Delta(T)\!=\!\Delta_0 \text{Tanh}(1.74 \sqrt{T_c/T - 1})$ where $\Delta_0$ is the pair potential at temperature $T=0$ and $T_c$ is the critical temperature of the superconductor. Throughout the present article (unless specified), $\mc{L}_{11}$, $\mc{L}_{12}$, and $\mc{L}_{22}$ are expressed in units of $e^2/h$, $k_Be/h$, and $k_B^2T/h$, respectively, where, $e$, $h$, and $k_B$ are the electron charge, Planck's constant and Boltzmann constant, respectively. Also, the natural units are considered where $\hbar=1$, $e=1$, $v_F\!=\!1$. All the energies are scaled by the zero-temperature pair potential ($\Delta_0$) and temperature is scaled by the superconducting critical temperature $T_c$. The superconducting coherence length is given by $\xi\!=\!\hbar v_F/\Delta_0\!=1$ when $\Delta_0\!=\!1$. The checmical potentials for all the normal regions are considered as $\mu_{\rm N}=0$ and for all superconducting regions $\mu_S=2$ unless specified.

\begin{figure}
\centering
\includegraphics[scale=.23]{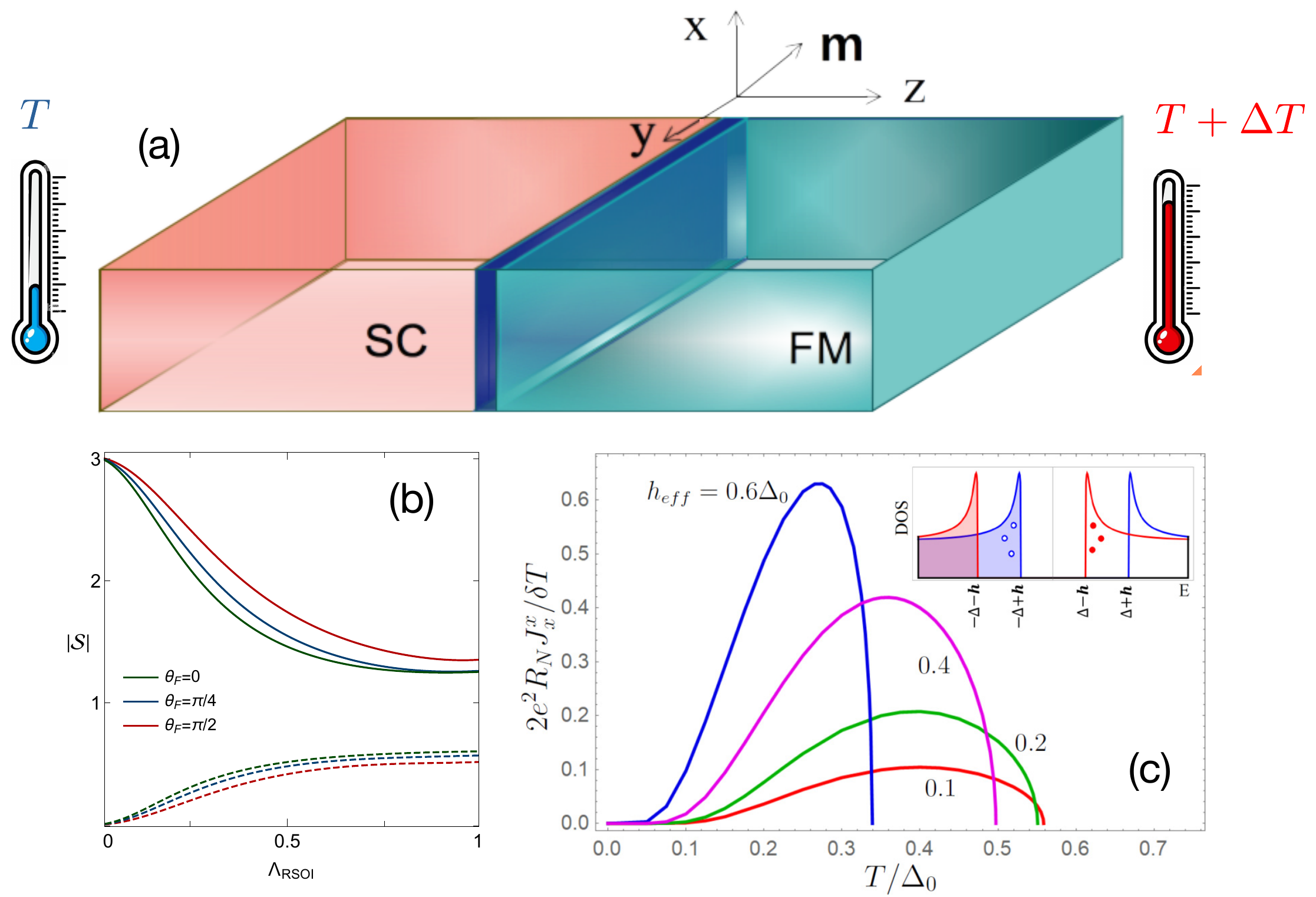}
\caption{(a) Schematic representation of FM/superconductor bilayer junction driven by temperature gradient and (b) Seebeck coefficient  (in units of $k_B/e$) for polarization $P=1$  of the FM as a function of the spin-orbit interaction separating the subgap contribution (dashed lines) from the total (solid) for various the polar angle $\theta_{\rm F}$ and the azimuthal angle $\phi_{\rm F}=0$ of the polarization vector of the FM. (c) Spin current per unit temperature gradient for different exchange field $h_{\rm eff}$. Inset: The spin-split density of states for the up and down spins are shown by red and blue color, respectively. The model and results are taken from (a-b) Ref.~\cite{Dutta2020b} and (c) Ref.~\cite{Bobkova2021}.  Copyright (2025) by the American Physical Society.}
\label{fig:FS}
\end{figure}

\subsection{Ferromagnet/superconductor bilayer} \label{Sec:FM}
Ferromagnetic layers have been used  in various forms in the superconductor junctions to obtain the giant thermopower (up to $\sim 100 \mu$V/K)~\cite{Ozaeta2014,Kolenda2016a}. The spin-splitting band structures of a FM (see inset of Fig.\ref{fig:FS}(c)) has been utilized to make the Andreev reflection process spin-dependent which eventually modifies the charge current in a thermally biased junction~\cite{Machon2014,Linder2016,Dutta2017}. In recent works on the FM/superconductor bilayer junctions, attention has been paid either to generate triplet Cooper pairs~\cite{Keidel2020,Dutta2020b} or to explore the possibility of advanced applications, particularly, spin-caloritronics applications~\cite{Bobkova2021,Tuero2025}. A very large thermal rectification is also achieved in ferromagnetic insulator based superconducting tunnel junctions~\cite{Giazotto2020}. 

A FM/ordinary superconductor bilayer junction shown in Fig.~\ref{fig:FS}(a) can be modelled by Bogoliubov-deGennes (BdG) equation as~\cite{Dutta2017,Dutta2020b},
\bea
\mc{H}_{\rm FS} (\mb{k})&=& \xi (k)\sigma_0 \eta_z-(\Delta_{xc}/2) \Theta(z) \mathbf{m}.\hat{\mathbf{\sigma}}\,\eta_z+\hat{H}_{int}\,\sigma_0 \eta_z \nonumber \\
&& ~~~~~~~~~~~~~~~~~~~~~~~~~~~~ + \Delta \Theta(z)\hat{\sigma_0} i \eta_y
\label{Eq:FS}
\eea
considering the interfacial barrier $\hat{H}_{int}=(V d\hat{\sigma_0}+ \mathbf{\omega}\cdot\hat{\sigma})\delta(z)$ with a $\delta$-function potential of height $V$, width $d$, and strength $Z$. A Rashba spin-orbit field described by $\omega$ $=\lambda[k_y,-k_x,0]$ and the interaction strength $\lambda_{rso}$ is also considered in Ref.~\cite{Dutta2020b}. The exchange spin splitting $\Delta_{xc}$ and the magnetization vector $\mathbf{m}=[\sin{\theta}\cos{\phi},\sin{\theta}\sin{\phi},\cos{\theta}]$ describes the FM. Here, $\xi(k)=k^2/2m$ is the kinetic energy of electrons with the effective mass $m$ measured from the chemical potential $\mu$ (set as: $m=1$, $\mu=0$), $\hat{\sigma}$ and $\eta$ are the Pauli spin matrices for the spin and Nambu basis, and $\Theta(z)$ is the Heavyside-step function. The superconducting pair potential of an ordinary $s$-wave spin-singlet superconductor is temperature-dependent as mentioned in the previous subsection.

In Ref.~\cite{Dutta2020b}, the role of unconventional spin-triplet odd-frequency Cooper pairs in the thermoelectricity
is studied in a FM/superconductor junction in the presence of a spin-active interface using the model Hamiltonian of Eq.\eqref{Eq:FS}. Enhanced thermopower is found in this FM/superconductor bilayer when odd-frequency spin-triplet Cooper pairs dominates over the conventional spin-singlet Cooper pairs. For the confirmation, the subgap contributions are also extracted as shown in Fig.~\ref{fig:FS} (b). The efficiency of the bilayer is also investigated in terms of the thermoelectric figure of merit and found to be large in the half-metallic limit of the FM. 

Later, the generation of equal-spin triplet Cooper pairs and the giant thermoelectric effect in the FM/superconductor bilayer junctions have been utilized to achieve spin-transfer torque effect in it. The spin-dependent electron-hole asymmetry around the Fermi energy results in large spin-Seebeck effect shown in Fig.~\ref{fig:FS}(c). It is sensitive to the temperature of the junction as well as the exchange field~\cite{Bobkova2021}. In the presence of a domain wall, this combination of the unconventional Cooper pair and large thermoelectric effect gives rise to thermally induced spin-transfer torques in FM/superconductor bilayer junction~\cite{Bobkova2021}. This type of magnetic control over the thermoelectric effects, being the central part of the spin-caloritronics, have become very popular now-a-days because of the potential of applications.

\subsection{Antiferromagnet/superconductor bilayer} \label{Sec:AFM}

To study the thermal transport in superconductor bilayers, AFMs are also added to the list of materials in 2020~\cite{Jakobsen2020}. The advantage of zero net magnetization inspite of magnetic ordering in AFM has been taken  to manipulate the scattering processes in these junctions. The vanishing stray field of the AFM has made it promising candidates for novel high-density and spintronic-based devices~\cite{Feldbacher2003,Demler2004,Kaczmarczyk2011}.  

The transport problem in the AFM/superconductor bilayer is solved using scattering matrix in Ref.~\cite{Jakobsen2020} using the model Hamiltonian given by,
\bea
\mc{H}_{\rm AFM} (\bf k) &=& \xi (k) s_z \tau_x \sigma_0 - \mu s_z \tau_0 \sigma_0 + J_0 s_4^-  \tau_z  (\mathbf{n}. \sigma)
+ V s_z \tau_0 \sigma_0 \nonumber \\ 
&& +\lambda_R s_4^+ \tau_x [(\sigma \times \mathbf{k}).\hat{z}] +\Delta(T) s_+ \tau_0 i\sigma_y  + \text{h.c.}
\label{Eq:H_AFM}
\eea
where $\xi(k)$ is the kinetic energy and $\mu$ is the chemical potential as mentioned in the previous subsections, $J_0$ is the exchange interaction strength. In the absence of this exchange interaction, the AFM behaves like a NM. $\Theta$ being the Heaviside step function, $\mathbf{n}=\{\sin \theta \cos \phi, \sin \theta \sin \phi, \cos \theta\}$ is the uniform Neel vector,  $V=V_0 \delta(z)$ is the spin-independent potential barrier, and $\lambda_R = \lambda_0 \delta(z)$ is the Rashba spin-orbit interaction term with the strength $\lambda_0$\,\cite{Jakobsen2020}. The superconducting term $\Delta(T)$ is taken as the $s$-wave spin-singlet term mentioned previously and it is set to zero in the normal part of the AFM junction. Here, $\tau$, $\sigma$, and $s$ are the Pauli matrices for the spin, sublattice, and charge degrees of freedom, respectively, $s^{\pm}_4=diag(1,\pm K)$, with $K$ is the complex conjugation, and $s^{\pm} = (s_x \pm i s_y )/2$.

\begin{figure}
\centering
\includegraphics[scale=.22]{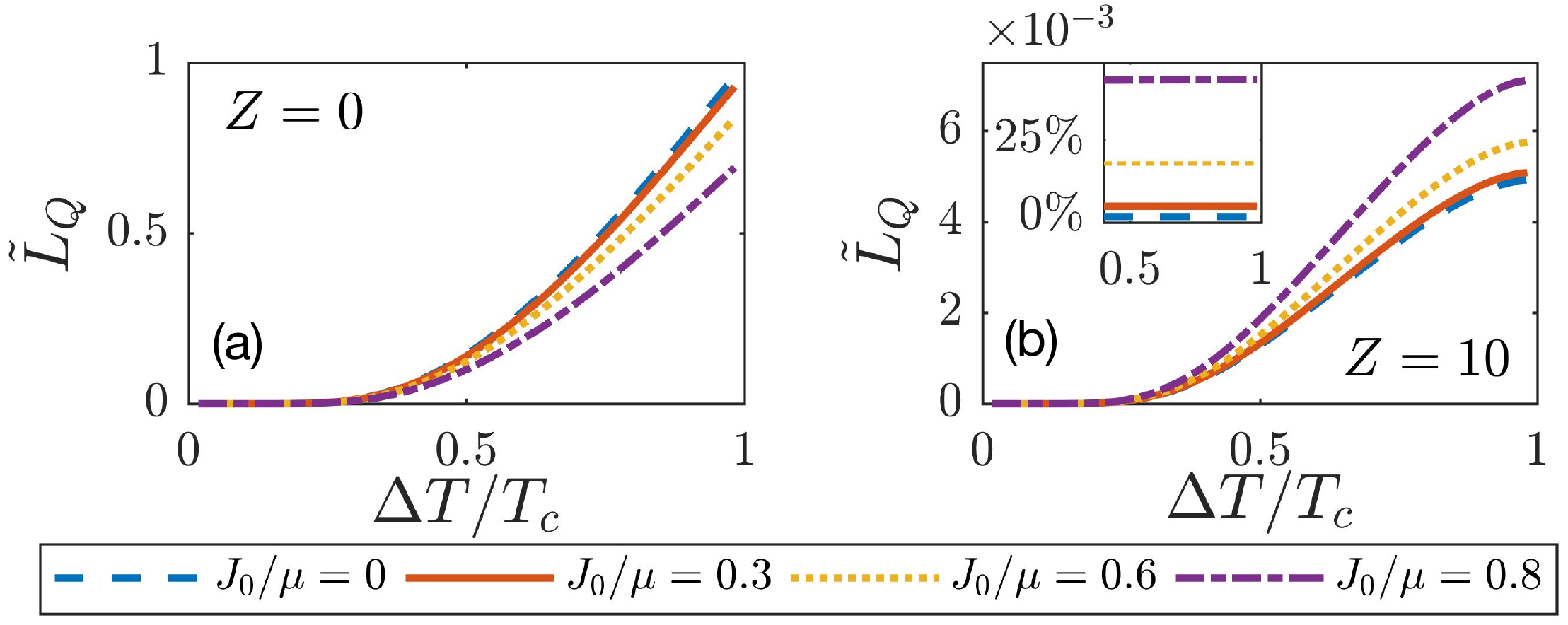}
\caption{Thermal conductance as a function of the temperature bias $\Delta T/T_c$ for different spin-independent barrier strength $Z$ and exchange strengths $J_0$. The insets depicts the peak in the percentage increase in the conductance as a function of the exchange interaction strength. All results are taken from Ref.~\cite{Jakobsen2020}. Copyright (2025) by the American Physical Society.}
\label{fig:results_AFM}
\end{figure}

The specular reflection of holes and retro reflection of electrons are the two scattering processes emerging at the AFM/superconductor interface and they determine the transport properties of this bilayer junction~\cite{Jakobsen2020}. The behavior of the thermal conductance in AFM/superconductor junction as a function of temperature is shown in Fig.~\ref{fig:results_AFM} for various exchange interactions and two limits of the barrier strength $Z = V_0 m/(\hbar^2 k_F^2)$ with $k_F^2=2m\mu/\hbar^2$~ \cite{Jakobsen2020}. Note that, the normalized thermal conductance is given by $\tilde{L}_Q=\int d^2 k_{||} \mc{L}_{22}/L_{Sh}$ where $k_{||}=\{ k_x,k_y,0\}$ is wave vector parallel to the interface and $L_{Sh}=A k^2_B T_c k_F^2/(12\hbar)$, with the interfacial area $A$, is the Sharvin thermal conductance evaluated for $\Delta_0 = J_0 = Z = 0$, resembling the perfect transmission for a NM. Note that, $\mc{L}_{22}$ is the same as mentioned in Eq.~\eqref{Eqn. L22}.

In Fig.~\ref{fig:results_AFM}, we present a few results from Ref.~\cite{Jakobsen2020} where the behavior of the thermal conductance is shown for the two limits of $Z$. In the transparent junction ($Z=0$), there is $100\%$ Andreev reflection which is of retro-type. The thermal conductance in this regime is suppressed since Cooper pairs do not carry any heat across the junction, but the finite charge. Thus, thermal conductance increases in the high temperature limit where the quasiparticles take part in the heat transfer process giving rise to the monotonic behavior of the thermal conductance. It can be decreased by tuning the exchange interaction strength since the retro normal reflection increases allowing less number of particles being transmitted into the superconductor side.
In the tunneling limit ($Z=10$), the number of quasiparticles transmitted into the superconductor layer decreases and the thermal conductance is strongly suppressed as we see in Fig.~\ref{fig:results_AFM}. However, interestingly, the thermal conductance increases with the increase in the exchange interaction strength which is in contrast to the transparent limit. This happens because of the effect of the interplay between the barrier strength and the exchange interaction on the quasiparticles. 
Thus, the behavior of the heat current with the exchange interaction is unique compared to the FM (or NM)/superconductor bilayer counterpart. A detailed comparison has been made in Ref.~\cite{Jakobsen2020}.

\subsection{Altermagnet/superconductor bilayer} \label{Sec:AM}

To the list of potential candidates for the thermal transport or thermoelectricity in the superconductor bilayers, recently discovered AMs have also been added in 2024~\cite{Chourasia2024,Sukhachov2024}. Altermagnetism is a form of magnetism with momentum-dependent spin-splitting originating from the underlying lattice geometry and the spin configuration~\cite{Smejkal2020,Yuan2020,Mazin2021,Smejkal2022a,Smejkal2022b,Mazin2022,Feng2022,Lee2024,Krempasky2024,Krempasky2024,Zhu2024}. The combination of the two lattice and the spin configuration leads to the simultaneous breaking of the parity and the time-reversal symmetry unlike spin-orbit coupled materials and lifts the spin-degeneracy. There are several material candidates for AMs ranging from metals to semiconductors and insulators like RuO$_2$~\cite{Ahn2019,Smejkal2020,Feng2022}, MnTe~\cite{Lee2024,Krempasky2024}, MnF$_2$~\cite{Yuan2020}, CrSb~\cite{Smejkal2022b}, FeSb$_2$~\cite{Mazin2021},  etc. The unique electronic structure of AMs makes them potential candidates for spintronic device applications and also provides pronounced crystal thermal transport due to Berry curvature in the momentum space~\cite{Zhou2024}. The absence of the magnetization in this class of materials leads to their insensitivity to the external magnetic field and thus, provides an advantage over the other bilayers based on the magnetic materials, where an unavoidable stray magnetic field persists and affects the results.
\begin{figure}
\centering
\includegraphics[scale=.22]{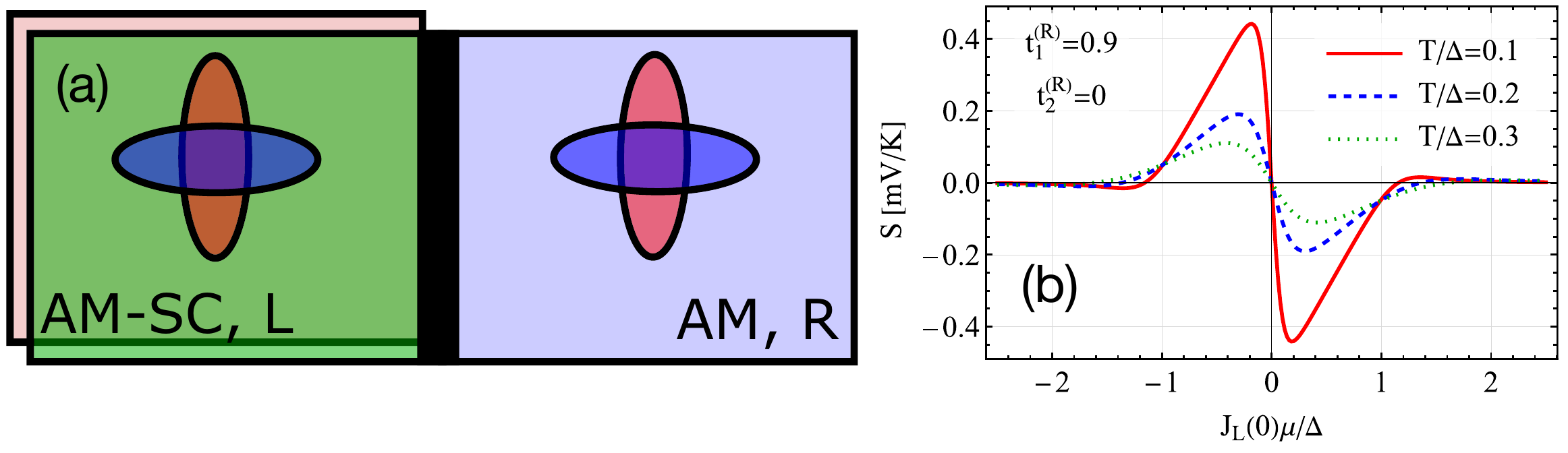}
\caption{(a) Schematic representation of $d$-wave AM/superconductor bilayer structure. (b) Seebeck coefficient as a function of the altermagnetic strength $J_L$ at $\varphi=0$ for various temperatures $T$. The model and results are taken from Ref.~\cite{Sukhachov2024}. Copyright (2025) by the American Physical Society.}
\label{fig:AFM}
\end{figure}

The BdG Hamiltonian to model the two-dimensional ($2$D) $d$-wave AM is used as~\cite{Sukhachov2024}, 
\bea
\mc{H}_{\rm AM}(\mbf{k})&=& \xi (k) +\sigma_z [t_1 (kx^2-k_y^2)/2] + t_2 k_x k_y +\Delta i \sigma_y \nonumber \\ 
&=& \xi (k) + \sigma_z ( \xi (k) + \mu) J_{\rm AM} +\Delta(T) i \sigma_y (\varphi)
\eea
where the kinetic enegy $\xi (k)$ as in the previous subsection, the parameters $t_1$ and $t_2$ are two dimensionless parameters describing the orientation and strength of the altermagnetic spin splitting\,\cite{Sukhachov2024}. The altermagnetic strength is defined in terms of $J_{\rm AM} (\varphi)=t_1 \cos{2\varphi} + t_2 \sin{2\varphi}$. The superconducting $\Delta(T)$ part is taken as zero in the normal part of the altermagnetic junction. A sizable thermopower has been predicted in the altermagnetic junctions driven by temperature gradient as shown in Fig.~\ref{fig:AFM}~\cite{Sukhachov2024}. Bilayer formed with AM and superconductor has been considered for the study of thermodynamics in Ref.~\cite{Chourasia2024} using quasiclassical Green's function. The specific heat found in this bilayer is different from the FM junctions and the difference lies into the absence of the first-order phase transitions for higher exchange fields which is found in the FM case. The study has been extended to check the response of the AM junctions to the external magnetic field and found anisotropic spin-susceptibility with one component similar to the bare superconductor revealing the nature of the $d$-wave nature of the AMs. Also, the inverse proximity effect in the superconductor by attaching AM and the spin-selective tunneling obtained by using another AM on the other side leads to reasonable Seebeck effect and the figure of merit as shown in Ref.~\cite{Sukhachov2024} using an effective model and functional integral approach. 

\subsection{Other material/superconductor bilayer} \label{Sec:OM}
Apart from the form of magnetism, other non-magnetic materials like Dirac materials e.g. graphene~\cite{Yokoyama2008,Paul2016}, silicene~\cite{Guzman2018} have also been used to study the thermal transport. In recent times, the tradition is continued to add bilayer graphene~\cite{Bernazzani2023,Bera2025} to explore the thermoelectricity using the unique properties of these materials. In addition to the response found in other Dirac fermions, some interesting features are found as an effect of the interplay between the trigonal warping and valley polarization. 
The Wiedemann-Franz (WF) law is violated near the charge-neutrality point, indicating the signatures of slow Dirac fermions~\cite{Bera2025}. Most interestingly, the opening of band-gap allowed enhanced Seebeck coefficient up to $1$\,mV/K, which is larger compared to ferromagnetic junctions~\cite{Bernazzani2023}. The thermoelectric figure of merit is also enhanced indicating the potential for applications.

Another promising route to induce the thermal current in superconducting interfaces with more advanced material like WSM~\cite{Saxena2023} is added to capture the interplay of the topology and superconductivity via thermal currents. WSMs in combinations with superconductors have been shown as a good candidate material to unravel intricate thermal transport properties. Specifically, inversion-symmetry-broken WSMs have been studied to form the superconducting bilayer junction using BTK formalism in Ref.~\cite{Saxena2023}. The thermal transport properties in this junction has been found to be sensitive to the doping. For low doping, the behavior is very similar to that of the NM, while for higher doping, it shows some unique feature like linearity in the thermal conductance which is contrast to the oscillatory behavior found in Dirc materials. The Lorentz number shows good matching with the metallic behavior for higher doping violating the WF law for small temperatures and low doping. Most interestingly, the figure of merit shows a sharp increase near the Weyl node, while it stays close to unity away from the Weyl nodes, and a sign-changing behavior of the thermopower is found indicating the change in the carrier with the increase in the barrier strength~\cite{Saxena2023}.

\subsection{Bilayer with unconventional superconductors} \label{Sec:BL}

Till now, we have discussed thermoelectricity in bilayer junctions where the superconductors are always ordinary $s$-wave spin-singlet type. In parallel to the replacement of the non-superconducting part of the bilayer junctions by advanced materials, various superconductors are also used to explore thermoelectricity in it. In the present subsection, we present the advancement in this direction. 

Thermoelectric response has been shown to offer a sensitive method to identify unconventional superconductors where pairing can happen via anisotropic channels~\cite{Savander2020,Guarcello2023,Matsushita2025}, particularly when the current is entirely due to the Andreev processes~\cite{Savander2020}. The possibilities of getting unconventional pairing which are different from the conventional $s$-wave spin-singlet pairing, encode their signatures in the Andreev dominated thermoelectricity. Specifically, surface Andreev states can be detected using a FM, FM insulator, and a Zeeman field as predicted by Savander \etal\,\cite{Savander2020}. The sign change of the electric current in a thermally biased junction has been predicted to appear due to the surface ABSs of unconventional superconductivity and thus carry the signatures of these surface states\,\cite{Savander2020}. The ABSs emerged at the NM/$d$-wave superconductor interface play a crucial role in the thermoelectric response\,\cite{Pal2024b}. The anisotropic nature of the $d$-wave pairing and the spontaneously generated topological Bogoloubov fermi surfaces (BFSs)~\cite{Pal2024a,Pal2024b} are helpful in that. 
\begin{figure}
\includegraphics[width=0.43\textwidth]{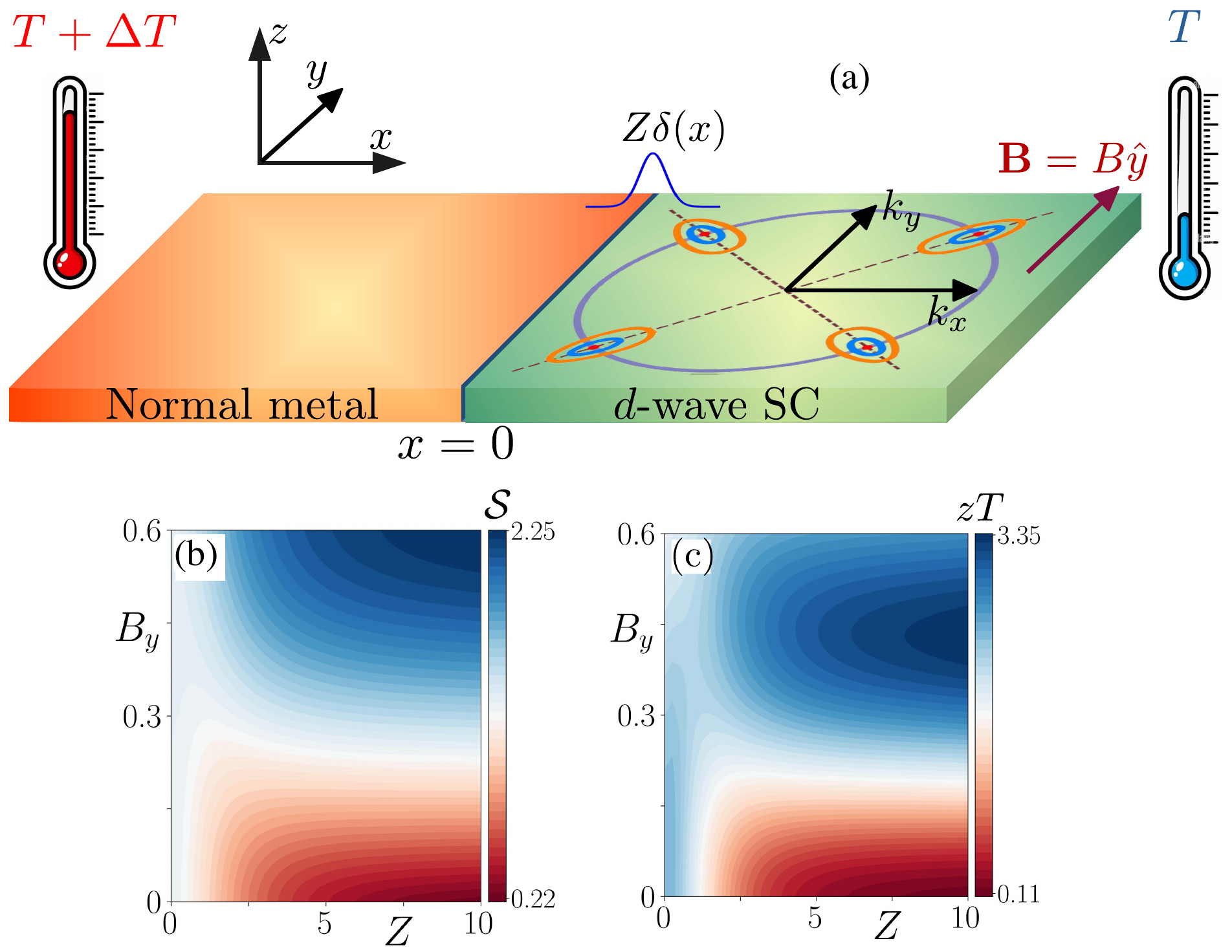}
\caption{(a) Schematic representation of NM/ $d$-wave superconductor bilayer with an interfacial $\delta$-function barrier of strength $Z$, an applied magnetic field ($\mbf{B}=B \hat{y}$) and a temperature gradient $\Delta T$ across the junction. Point nodes (red dots), line nodes (dashed lines) of bare $d$-wave superconductor, and Bogoliubov Fermi surfaces (blue and orange elliptic contours) are shown along with the NM fermi surface (large grey circle). (b) Seebeck coefficient (measured in units of $k_B/e$) and (c) figure of merit, $zT$ in the $B_y\!-Z$ plane for $T/T_c=0.2$ and $\alpha=\pi/4$. The model and results are taken from Ref.~\cite{Pal2024b}. Copyright (2025) by the American Physical Society.} 
	\label{Fig:BFS}
\end{figure}

The model Hamiltonian to describe a $2$D planar NM/ $d$-wave superconductor bilayer with an in-plane magnetic field $\bf{B}$ applied to the superconductor region (shown in Fig.~\ref{Fig:BFS}(a)) is given by~\cite{Setty2020PRB,Pal2024b}
\begin{equation}
\mc{H}_{\rm BFS}(\mb{k})=\xi (k) \tau_z \sigma_0 +\Delta (k,\alpha ) \tau_x \sigma_0 - B \tau_0 \sigma_y
\end{equation} 
where the kinetic energy part is given by that of a $2$D electron gas as mentioned in the previous subsections and the $d$-wave superconducting pair potential is in the form $\Delta (k, \alpha )=\Delta_0 \cos [2 (\theta + \alpha )]$ with $\theta=\tan ^{-1}(k_y/k_x )$. Here, $\alpha$ denotes the angle between the $a$-axis of the crystal and normal to the interface\,\cite{Yang1998,Pal2024b}. In Ref.~\cite{Pal2024b}, a detailed investigation has been made in such a model to explore the signatures of the unconventional BFSs generated in such a junction with $d$-wave superconductor in terms of the thermal conductance, Seebeck coefficient, figure of merit, and the WF law. There appear ABSs in the interface with the change in the orientation angle of the $d$-wave vector, $\alpha$ from zero to $\pi/4$. It gives rise to a significant enhancement in the Seebeck coefficient due to the generation of BFSs. The maximum value of $S\sim 2.25k_B/e \sim 200 \,\mu V/K$ at $T/T_c=0.2$ as depicted in Fig.~\ref{Fig:BFS}(b). This value of the thermopower is significantly larger than the thermopower found in ferromagnetic junctions ($\sim 100 \mu V/K$)~\cite{Kolenda2016a}. An external gate voltage at the junction is also applied which can tune the junction transparency. This enhancement of the Seebeck coefficient eventually results in a large figure of merit i.e., higher rate of conversion from the heat energy to the electric current indicating higher thermoelectricity using a $d$-wave SC in presence of a magnetic field hosting BFSs. It is higher compared to other $d$-wave superconductor junctions without any magnetic field~\cite{Devyatov2000,Yokoyama2005} and also higher than the thermopower found bare $d$-wave superconductor ($\sim$ a few $\mu$V/K)~\cite{Seja2022}. A maximum value of $zT\sim 3.5$ is observed in this heterostructure which carry signatures of unconventional superconductor with BFSs and on top of that, proves it as a potential candidate for thermoelectric applications. To note, in addition to the local thermoelectric effects, nonlocal thermoelectric effects have also been shown to be generated in superconductors with BFSs and predicted as signatures of BFSs~\cite{Mateos2024}.
Furthermore, superconducting tunnel junctions between Fe-based superconductor and BCS superconductor is predicted to show very large thermoelectric effect with Seebeck coefficient $\sim 800 \mu$ V/K and figure of merit $zT > 6$ at a few Kelvin~\cite{Guarcello2023}. 

\section{Josephson junction}\label{Sec:JJ}

Now, we present a discussion on the progress in the study of the thermal transport in three-layered heterostructures among which superconductor Josephson junction is the most widely known and useful junction. The tunability of the superconducting phase-difference in Josephson junction provides an additonal freedom which leads to the phase-tunable thermal current in such junctions~\cite{Blasi2020a,Blasi2020b,Dutta2023}. Similar to the bilayer junctions, several materials have been considered in Josephson junctions too. Some recent advancement in the Josephson junctions along with the theoretical formalism are summarized as follows.

\subsection{Theoretical formalism}
Within the linear regime, the charge current can be expressed using the Landauer formalism as 
\begin{align}
I^c &= \frac{2e}{h} \Delta T \int_0^\infty d\omega \left[  i^{e}_{\rm L}(\omega)  -  i^{h}_{\rm L}(\omega) \right] \frac{\partial f\left(\omega/T\right)}{\partial T}
\label{Eq:Ic}
\end{align}
where the net current is found from the difference between the probabilities of transmission of quasiaprticles along the opposite directions given by\,\cite{Dutta2017,Pershoguba2019}
\begin{align}
 i_L^\eta=T_{\eta\eta}^{RL} - T_{\eta^{\prime}\eta}^{RL}.
 \label{eq:ie}
\end{align}
with $\eta \in {\rm \{e, h\}}$ and $T_{\eta\eta}^{l^{\prime}l}=| t_{\eta\eta}^{l^{\prime}l} |^2$, where $T^{l^{\prime}l}_{\eta^{\prime}\eta}$ ($t^{l^{\prime}l}_{\eta^{\prime}\eta}$) is the probability (amplitude) of the transmission of $\eta^{\prime}$ type particles from $l^{\prime}$-th to the $l$-th lead as $\eta$. Here, $e$ is the electronic charge, $h$ is the Planck's constant, $\omega$ is the incoming electron energy, $f$ is the Fermi distribution function, $i^{e(h)}_L$ denotes the contributions by the electrons (holes) in the left lead accordingly. To be noted, an usual non-dissipative Josephson current flow is nonvanishing even at $\Delta T=0$ when $\Delta \phi\ne 0$. The expressions for the transmission amplitudes can be found using scattering matrix or Green's function formalism. The lower limit of the integration in Eq.\,\eqref{Eq:Ic} can be replaced by the superconducting gap unless $T_{\rm ee}^{\rm R}\!=0$ within the gap.
\begin{figure}
\centering
\includegraphics[scale=.25]{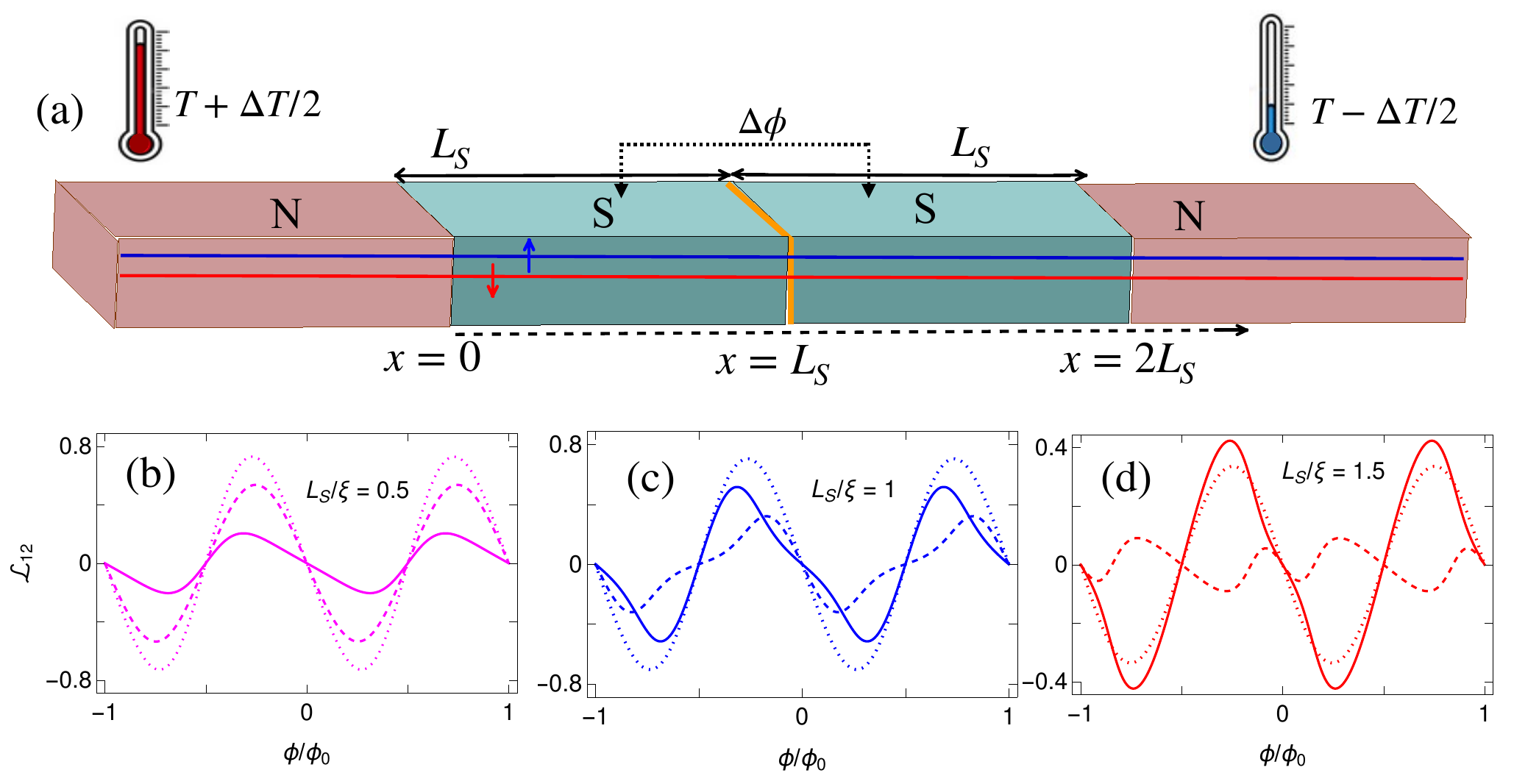}
\caption{(a) Schematic representation of topological Josephson junction, formed at the edge states (up and down spin channels are marked by red and blue, respectively) of $2$D topological insulator using ordinary $s$-wave superconductors, biased by temperature gradient. (b) Charge current per unit temperature gradient $\mathcal{L}_{12}$ (in units of $ek_B/h$) as a function of $\phi/\phi_0$  for various system size $L_{\rm S}$. The solid, dashed, and dotted lines denote three integration limits $\{0,\Delta_0\}$, $\{\Delta_0,\infty\}$, and $\{0,\infty\}$, respectively, The model and results are taken from Ref.\cite{Dutta2023}.}
\label{Fig:TopoJJ}
\end{figure}

The non-dissipative charge current per unit temperature gradient is defined as the thermopower or the Seebeck coefficient 
which provides the amount of the charge current produced due to the thermal gradient. The heat current induced by the temperature gradient is expressed by
\begin{align}
I^q &= \frac{2}{h} \Delta T \int_0^\infty d\omega \left[  i^{e}_{\rm L}(\omega)  +  i^{h}_{\rm L}(\omega) \right] \frac{\partial f\left(\omega/T_l\right)}{\partial T}. 
\label{Iq}
\end{align}
To note, this charge current per unit temperature gradient is not the conventional Seebeck effect as we explain in the next section. Reversing the phase can help in separating the non-dissipative charge current from the dissipative current\,\cite{Pethick1979,Clarke1979,Pershoguba2019}.

Nonlinear thermoelectricity of a Josephson junction has also drawn the attention~\cite{Bours2019,Marchegiani2020}. For the nonlinear regime, higher order terms of the Fermi distribution function is considered. 

\subsection{Topological Josephson junctions}\label{Sec:Topo}

Similar to bilayer junctions, topology has been added to Josephson junction too in recent years. Attention has been paid to the utilization of the topological quantum matter to enhance the thermoelectric effects on one side\,\cite{Germanese2022,Marchegiani2020,Martinez2013,Chatterjee2024}. On the other hand, the phase-dependent heat current has been used as an alternative way to probe topological features\,\cite{Sothmann2016a}. Among various junctions, Josephson junctions based on topological insulators have attracted a lot of interests because of its intriguing properties and the potential to host exotic states~\,\cite{Kane2005,Fu2008,Fu2009,Qi2011,Tkachov2013,Hwang2020}. We now discuss the progress in topological Josephson junction in the present context. 

The BdG Hamiltonian describing the topological Josephson junction shown in Fig.~\ref{Fig:TopoJJ}(a) is considered as~\cite{Dutta2023},
\bea
\mc{H}_{\rm TIJJ} (\mb k)=-i v_{F}\partial_{x}\tau_z\sigma_{z}-\mu\tau_z\sigma_0 +\Delta (x) \tau_x i\sigma_y
\label{Eq:Ham_TJJ}
\eea
where the kinetic energy of electrons describes the linear dispersion relation of the edge states of $2$D topological insulator with the Fermi velocity $v_F$ and chemical potential $\mu$. The proximity-induced superconductivity takes the following values: $\Delta (x) =\Delta$ for the left superconductor ($0\leq x \leq L_{\rm S}$), $\Delta (x) =\Delta e^{i\phi}$ for the right superconductor ($L_{\rm S}\leq x \leq 2L_{\rm S}$), and $\Delta (x)= 0$ for all the normal parts ($x<0$ and $x> 2L_{\rm S}$). The Pauli matrices $\sigma_{i}$ and $\tau_i$ act in the spin and Nambu space, respectively. The phase difference between two superconductors of the thermally biased junction can be tuned externally by applying a magnetic flux $\phi$. 
In Ref.~\cite{Dutta2023}, all thermal transport coefficients are calculated based on two simplifications: (i) the middle normal region sandwiched between two superconductors is taken as tiny and (ii) lengths of the two superconductors are considered as equal without any loss of generality~\cite{Dutta2023}. 

The helical edge states of topological insulator allows allows only two processes: (i) Andreev reflections and (ii) electron transmissions through the junction prohibitting all backscatterings and this eventually modifies the transport coefficients~\cite{Hasan2010,Cayao2017,Cayao2022}. At the central region of the junction, topological ABSs protected by fermion parity appear indicating the appearance of Majorana zero modes when the superconducting phase difference is tuned to $\phi=\pi$~\cite{Cayao2022,Dutta2023}.  

The charge current having oscillatory profiles with $\phi/\phi_0$ as depicted in Fig.~\ref{Fig:TopoJJ}(b-d) satisfy symmetric condition: $\mathcal{L}_{12}(\phi/\phi_0)=-\mathcal{L}_{12}(-\phi/\phi_0)$ and $\mathcal{L}_{12} (\phi/\phi_0\!=\!n)=0$ where $n$ is an integer, indicating the reciprocity behavior of the current. Importantly, the absence of the even-symmetry part of the charge current indicates the absence of the thermoelectricity (dissipative current) in finite-size topological Josesphson junction for all phases~\cite{Dutta2023}. The odd-symmetric finite charge current appears due to the symmetry-breaking of the transmission probability around $\omega=0$ as discussed in Ref.~[\onlinecite{Dutta2023}]. Similar study has been made in Ref.~[\onlinecite{Mukhopadhyay2022}] where superconductors are taken as leads instead of finite sizes. The reason behind the absence of dissipative charge current has been attributed to the interference between the currents carried by Cooper pairs and quasiparticles~\cite{Mukhopadhyay2022}. 

Here appears a question about the contributions by the sub-gap and super-gap states to the total current and also the sensitivity of the current to the junction size. This has been analyzed by the breaking the integration limit of Eq.~\eqref{Eq:Ic} in two steps ($[0,\Delta_0]$, [$\Delta_0,\infty$]) separating into the sub-gap and super-gap limits (shown in Fig.~\ref{Fig:TopoJJ}(b-d)). The sub-gap and super-gap contributions are in phase giving rise to the additive total charge current in the limit $L_{\rm S}/\xi\ll 1$ with $\xi$ being the superconducting coherence length, whereas, they posses opposite phase giving rise to the lower total charge current for $L_{\rm S}/\xi> 1$. 
For $L_{\rm S}/\xi\ll 1$, the total charge current flowing through the junction is dominated by the quasiparticles above the superconducting gap. The contributions by the sub-gap states to the thermally induced total charge current is highest and in phase when $L_{\rm S}/\xi\sim 1$. Beyond this limit, the contributions by ABSs are majorly compensated by the contributions from the super-gap states. Although the symmetry-breaking around $\omega\!=\!0$ has been utilized to induce the charge current in the system, it is not an essential criterion for the generation of the heat current. 
A detailed analysis of the heat current is also made in Ref.~\cite{Dutta2023}. For the optimization of the efficiency of a thermoelectric current, short junction size is preferable at low temperature. 

Similar topological Josephson junction has been explored to study the heat capacity in Ref.~[\onlinecite{Scharf2021}]. An enhancement of heat capacity controlled by the phase difference is found even when the fermion parity is not conserved. In Ref.~\cite{Scharf2021}, the authors showed that the heat capacity is characterized by the appearance of a double peak while varying as a function of the phase difference. The double peak 
is predicted as a signature of the protected zero-energy crossing in the Andreev spectrum. For the trivial junction, there appears a gap around the zero-energy irrespective of the trivial ABSs. 
In the presence of a finite magnetic flux a Doppler shift of the Cooper pair momentum described by $p_s$ is induced. 

Topological Josephson junction has also been predicted to act as a passive thermal rectifier and thermal diode recently in Ref.\,[\onlinecite{Bours2019}], enriching its potential from the application perspective. The Doppler shift effect due to the interplay of the superconducting leads and helical edge states has been utilized to illustrate a rectification scheme in topological counterpart of the TSQUIPT~\cite{Bours2018,Bours2019}. It offers a control over the superconducting gap via a small magnetic field, thus regulating the heat flow through the junction. Both linear and nonlinear regime have been explored in Ref.~\cite{Bours2019}. In fact, topological Josephson junction with an additional normal metal probe is a suitable testbed to study the nonlocal thermoelectric effect~\cite{Blasi2020a,Blasi2020b}. Nonlocal thermoelectric effect is shown to carry a unique signature of the helical edge states in another work by Blasi \etal where the Doppler shift is controlled by an external magnetic flux~\cite{Blasi2020a}.

Thermal transport properties of topological Josephson junction can carry signatures of topological phases when magnetism is included~\cite{Gresta2021}. Depending on the number of the magnetic islands, there can appear Jackiw-Rebbi solitons and Majorana zero modes captured via the thermal conductance. Remarkably, in the case of solitons, the decreasing behavior of the thermal conductance with the temperature can be used to identify these modes, whereas the Josephson current is not sensitive to this resonant states\,\cite{Gresta2021}. The role of Majorana bound states in the context of the thermoelectric effect has been studied in other works too~\cite{Mukhopadhyay2021, Smirnov2025}. 

Thus, topological Josephson junctions have been studied in great detail with lots of possible applications which can be utilized in thermal devices based on superconductor.

\subsection{Weyl Josephson junction} \label{Sec:WSM}

\begin{figure}
\centering
\includegraphics[scale=.25]{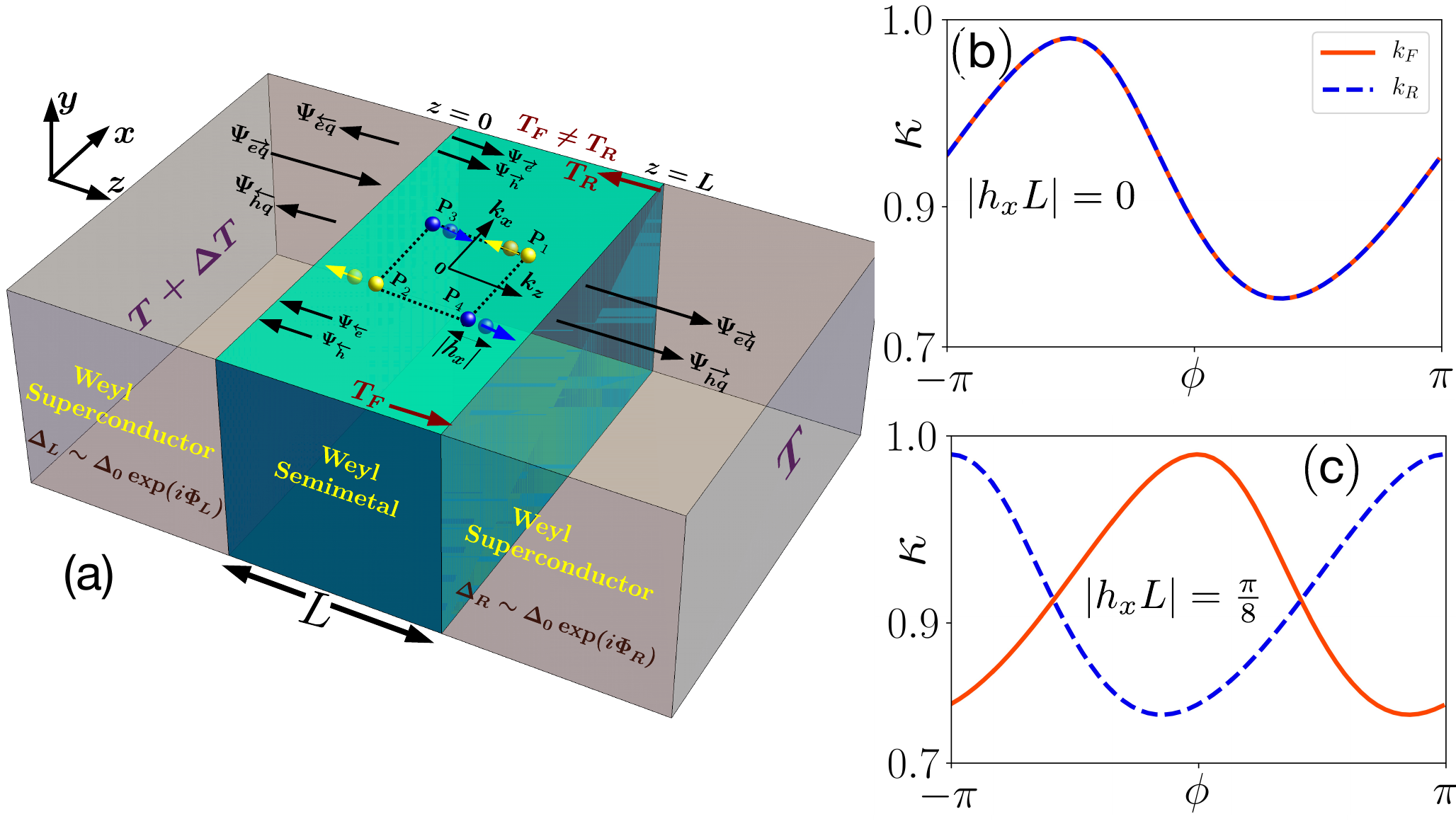}
\caption{(a) Schematic representation of WSM Josephson junction. (b-c) Thermal current (in units of $k_B^2/(2\pi h)$) along the forward and reverse direction in the (b) absence and (c) presence of magnetic field keeping $L=0.5$. The model and results are taken from Ref.\,\cite{Chatterjee2024}.}
\label{fig:WSM}
\end{figure}

Reciprocal relations in irreversible processes~\cite{Onsager1931a,Onsager1931b} are valid unless we break the symmetry. Effort has been put forward to explore the effect of the symmetry breaking. For this, other topological materials like WSMs, which is an explicit realizations of high energy phenomena in low-energy condensed matter systems, have been utilized. The appearance of the nonreciprocal current i.e., the so-called diode effect based on thermal current has been found in Weyl based Josephson junction shown in Fig.~\ref{fig:WSM}(a)~\cite{Chatterjee2024}.  

The low density of states at the gapless nodal points known as Weyl nodes originates from the symmetry breaking which essentially help in manipulating the heat current carried by quasiparticles in this junction. An asymmetry between the forward and reverse heat currents appears, which is tunable by the external magnetic field\,\cite{Chatterjee2024}. In fact, one of the two heat currents can be immensely suppressed by optimizing the junction length, magnetic field strength, and superconducting phase difference as shown in Fig.\,\ref{fig:WSM}(b-c). The external tunability along with high rectification greatly enhances the usefulness of the junction. 

An inversion-symmetry broken WSM Hamiltonian is given by\,\cite{Zhang2018a,Zhang2018b,Saxena2023}
\begin{align}
\mc{H}_{\rm WSM}(\mb{k}) & =k_{x}\eta_{x}\sigma_{z}+k_{y}\eta_{y}\sigma_{0}+(\kappa_{0}^{2}-|{\bf k}|^{2})\eta_{z}\sigma_{0}\nonumber \\
& \ \  ~~~~~~~~~~~~~ -\alpha k_{y}\eta_{x}\sigma_{y} +\beta \eta_{y}\sigma_{y},
\label{Eq1}
\end{align} 
with the two Pauli matrices $\eta$ and $\sigma$ acting on the orbital and spin degrees of freedom, respectively. The other parameters $\kappa_{0},$ $\alpha$ and $\beta$ are model-dependent parameters describing the WSM with four Weyl nodes  locating at P$_{1,2}=\pm\left(\beta,0,\sqrt{\kappa_{0}^{2}-\beta^{2}}\right)$ and P$_{3,4}=\pm\left(\beta,0,-\sqrt{\kappa_{0}^{2}-\beta^{2}}\right)$ assuming $0$$<$$\beta$$<$$\kappa_{0}$\,\cite{Zhang2018a,Zhang2018b}, where P$_1$ and P$_2$ (P$_3$ and P$_4$) carry the positive (negative) chirality. They are the time-reversed pairs. For the Josephson junction, the low-energy Hamiltonian has been extracted along with an applied magnetic field $h(\bf{r})$ as,
\beq
\mc{H}_{\rm WJJ}(\phi)=h(\mathbf{r})\sigma_z\nu_0-\mu(\mathbf{r})\sigma_0\nu_z-i\partial_{\mathbf{r}}\cdot\mathbf{\sigma}\nu_z+\Delta_s(\mathbf{r})\sigma_0\nu_x
\eeq
where the Pauli matrix $\nu$ acts on the particle-hole space\,\cite{Zhang2018a,Zhang2018b,Saxena2023} and $h(\mathbf{r})=h_x\Theta(z)+h_x\Theta(L-z)$. The pair potential is taken as finite in the two WSM superconductors and zero in the middle normal WSM region. In the absence of the magnetic field, the behavior of the thermal conductance along the forward and the reverse directions are identical to each other (see Fig.~\ref{fig:WSM}(b)). As soon as the magnetic field is applied, the asymmetry being sensitive to the magnetic field grows. This phenomenon has been utilized in Ref.\,[\onlinecite{Chatterjee2024}] to model thermal diode using WSM Josephson junction as shown in Fig.\,\ref{fig:WSM}(c).  

\subsection{Other Josephson junctions} \label{Sec:WSM}
\begin{figure}
\centering
\includegraphics[scale=.25]{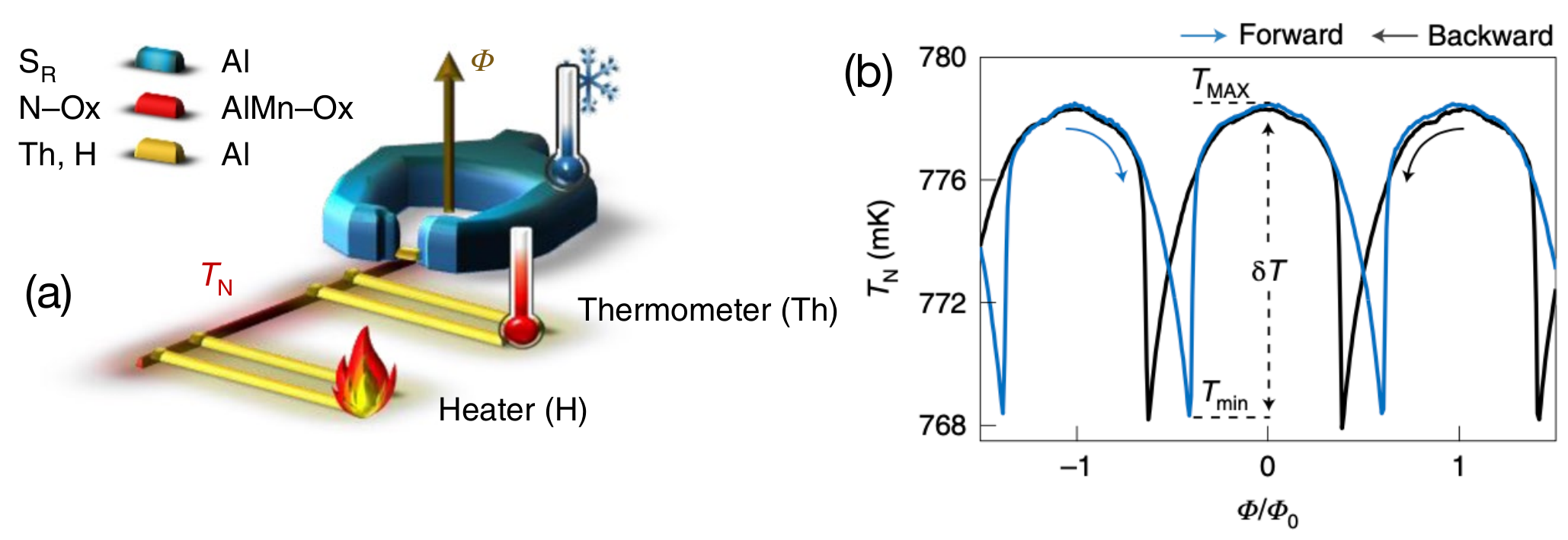}
\caption{(a) Schematic representation of the TSQUIPT. The electronic temperature $T_N(\phi)$ in the normal electrode (red color; Al$_{\rm 0.98}$Mn$_{\rm 0.02}$) is controlled by two pairs of superconductors (yellow; Al; served as heaters and coolers) tunnel coupled to the normal part. (b) The electronic temperature $T_N (\phi)$ acquired for $P_{\rm in}= 3.2$ pW. The blue (forward) and black (backward) curves show different magnetic-flux sweep directions. $T_{\rm max}$ and $T_{\rm min}$ of $T_{\rm N}$ along with the temperature swing ($\delta T$) are marked. The model and result are taken from Ref.\,\cite{Ligato2022}. Copyright (2025) by the
Springer Nature.}
\label{Fig:TSQUIPT}
\end{figure}

In Josephson junctions, one of the most remarkable works in recent times is the realization of the TSQUIPT in a quasi one-dimensional ($1$D) Al nanowire forming a weak-link embedded in a superconducting ring as shown in Fig.\ref{Fig:TSQUIPT}(a)~\cite{Ligato2022}. The possibility of controlling the heat current by phase tuning the superconducting proximity effect is a great advantage. It is achieved by the manipulation of the density of states by the magnetic flux~\cite{Ligato2022}. The temperature modulations achieved by the phase bias yields temperature-to-flux transfer function and a hysteretic dependence of the local density of states on the applied field due to phase-slip transitions. This  enables the TSQUIPT device to operate as a phase-tunable thermal memory encoding the information in the temperature of the metallic region illustrated in Fig.~\ref{Fig:TSQUIPT}(b).
Thermal noise effect is also studied in ferromagnetic Josephson junction~\cite{Guarcello2021}. Other Josephson junctions mostly include multi-terminal Josephson junctions. Thermally-biased Josephson junctions have been investigated to predict phase-sensitive current along with the formation of a $\pi$-junction controlled by the temperature gradient and the system topology, by Kalenkov \etal~\cite{Kalenkov2020}. They have demonstrated that long-range Josephson effect and shown that the temperature gradient induced voltage is dominated by the non-equilibrium low-energy quasiparticles at temperatures strongly exceeding the Thouless energy where equilibrium Josephson current negligibly contribute~\cite{Kalenkov2020}. Josephson junctions based on advanced materials include $2$D buckled material~\cite{Zhuo2023}. The buckling nature of the material has been used to have electrical control over the phase-dependent thermal transport in both antiferromagentic and ferromagnetic Josephson junctions based on buckled materials. Majorana-mediated thermoelectric transport in multiterminal Josephson junctions has been studied~\cite{Klees2024}. Very recently, phase-coherent thermal transport phenomena have been predicted in Josephson junctions based on $d$-wave AMs~\cite{Chen2025}.
In the context of the multilayers using unconventional superconductors, the thermoelectric transport in Josephson junctions have been studied to explore the role of Majorana zero modes or bound states hosted by topological superconductors~\cite{Klees2024,Trocha2025}

\section{Other multilayer heterostructures}\label{multi}

Apart from the Josephson junctions, other form of multilayer superconductor heterostructures have also kept their remarks in this context. They include AM/superconductor/AM junction to predict enhancement in the Seebeck coefficient and the thermoelectric figure of merit~\cite{Sukhachov2024}, ferromagnetic insulator/superconductor/insulator/ferromagnet tunnel junction to explore superconducting spintronic heat engine~\cite{Araujo2024}, thermally biased SIS$^{\prime}$IS junction using the combination of materials EuS/Al/AlOx/Co to show the bipolar thermoelectric superconducting single-electron transistor operation~\cite{Battisti2024}. A complete control over the thermoelectric effect by magnetic field has been obtained in 
weak FM/fully spin-polarized ferromagnetic insulator when coupled to a superconducting reservoir (S) and NM reservoir on two sides~\cite{Ouassou2022}.

The progress in experiments includes the realization of superconducting spintronic heat engine in a ferromagnetic insulator/superconductor/insulator/ferromagnet tunnel junction that converts thermal energy into an usable energy~\cite{Araujo2024}. The sign of the thermoelectric voltage generated in this junction can be changed by changing the orientation of the two ferromagentic layers placed at the two ends of the junction. A large thermoelectric figure of merit has been explored in chiral superconductor junction where the junction is formed by arranging quantum anomalous hall insulator/topological superconductor/quantum anomalous hall insulator~\cite{Chen2025} Very recently, large thermoelectric spin-valve effect is realized in FM/superconductor/FM junction and the Seebeck coefficient ($\sim 100 \mu$V/K) found in this junction shows a very good agreement with the previous works in FM/superconductor bilayer junction~\cite{Tuero2025}. 

\section{Summary and conclusions} \label{Sec:Summary}

In this article, we have reviewed recent advancements during the last five years in the two-terminal thermal transport in superconducting heterostructures. Among various superconducting junctions, ferromagnetism in combination with superconductors has drawn significant attention since the last decade due to the rich emergent phenomena in these systems. In addition to discussions on recent developments in the thermal transport phenomena in these ferromagnet/superconductor junctions, we have summarized results in junctions where ferromagnets are replaced by other forms of magnetism like antiferromagnetism and altermagnetism. In order to enhance the thermoelectricity, other topological materials have also been investigated. Major attention has been paid to Josephson junctions, particularly, topological Josephson junctions due to the possibility of the phase-tunable caloritronics opening a pathway for various applications. Along with the superconducting phase, the system size can also be a confirming factor for the thermoelectricity in them. However, the progress is not limited to bilayers or Josephson junctions; multilayer interfaces using superconductor(s) with various other nonsuperconducting materials have been reported recently. In parallel to ordinary superconductors, unconventional superconductors are also predicted to be useful in the present context. On top of that, thermometry has been utilized to detect topological or other bound states or emergent phases in the superconducting heterostructures. 

In most electronic devices, heat dissipation sets constraints on their performances; thus maintaining the device temperature is a key factor and becomes challenging. Management of heat and its conversion via thermoelectricity helps for various functionalities with potential applications in superconductor-based thermal devices such as sensitive electron thermometers~\cite{Giazotto2015}, thermoelectric radiation detector~\cite{Heikkila2018,Chakraborty2018}, TSQUIPT~\cite{Bours2018,Bours2019,Ligato2022}, magnetic switch~\cite{Guarcello2021}, thermoelectric bolometer and calorimeter with a superconducting quantum interference device readout~\cite{Geng2020}, superconducting spintronic heat engine~\cite{Araujo2024}, rectifiers~\cite{Martinez2015,Chatterjee2024}, thermally driven superconducting single-electron transistors, ~\cite{Sothmann2021}, logic gate operations using thermally biased Josephson junctions~\cite{Paolucci2018}, building block for superconducting computer memory~\cite{Shafraniuk2019} etc. The search for various configurations of superconducting heterostructures using traditional as well as advanced materials with newer functionalities and applications is continuing ~\cite{Jakobsen2020,Savander2020,Keidel2020,Gresta2021,Bobkov2021,Ligato2022,Bernazzani2023,Dutta2023,Pal2024b,Chatterjee2024,Araujo2024}. 

\begin{acknowledgments}
P.\,D. thanks Arijit Saha, S.\,D.\,Mahanti, Kau\^{e} R. Alves, Alhun Aydin, Altug Sisman, Jonas Fransson, Martin F.\,Jakobsen, Kristian B.\,Naess, Alireza Qaiumzadeh, Navinder Singh, Pritam Chatterjee, Amartya Pal, Debika Debnath, and Annica M.\,Black-Schaffer for all discussions on the present topic and acknowledges Department of Space (DoS), India for all support at Physical Research Laboratory. P.\,D. thanks Ranjan Laha for helpful discussions. I devote this article to the memory of late Prof.\,Arun M.\,Jayannavar.
\end{acknowledgments}

\bibliography{bibfile}

\end{document}